\documentclass[final,5p,times]{elsarticle}

 \usepackage{graphicx}

\usepackage{xspace}

\usepackage{float}

\newcommand{\kb}{\ensuremath{k_{b}}\xspace}
\newcommand{\fb}{\ensuremath{f_{b}}\xspace}
\newcommand{\fm}{\ensuremath{f_{m}}\xspace}
\newcommand{\km}{\ensuremath{k_{m}}\xspace}
\newcommand{\ka}{\ensuremath{k_{a}}\xspace}

\newcommand{\R}[1]{\ensuremath{R_{#1}}\xspace}

\newcommand{\wrf}{\ensuremath{\omega_{rf}}\xspace}
\newcommand{\wi}{\ensuremath{\omega_{1}}\xspace}
\newcommand{\wa}{\ensuremath{\omega_{a}}\xspace}
\newcommand{\wb}{\ensuremath{\omega_{b}}\xspace}
\newcommand{\dw}{\ensuremath{\Delta\omega}\xspace}
\newcommand{\dwb}{\ensuremath{\Delta\omega_{b}}\xspace}
\newcommand{\db}{\ensuremath{\delta_{b}}\xspace}

\newcommand{\B}{\ensuremath{B_{0}}\xspace}
\newcommand{\vB}{\ensuremath{\vec{B}_{0}}\xspace}
\newcommand{\bi}{\ensuremath{B_{1}}\xspace}
\newcommand{\A}{\ensuremath{\mathbf{A}}\xspace}
\newcommand{\As}{\ensuremath{\mathbf{A'}}\xspace}

\newcommand{\weff}{\ensuremath{\omega_{\mathit{eff}}}\xspace}
\newcommand{\vweff}{\ensuremath{\vec{\omega}_{\mathit{eff}}}\xspace}
\newcommand{\xeff}{\ensuremath{x_{\mathit{eff}}}\xspace}
\newcommand{\yeff}{\ensuremath{y_{\mathit{eff}}}\xspace}
\newcommand{\zeff}{\ensuremath{z_{\mathit{eff}}}\xspace}

\newcommand{\Mv}{\ensuremath{\vec{M}}\xspace}
\newcommand{\vmss}{\ensuremath{\vec{M}^{\mathit{ss}}}\xspace}
\newcommand{\mss}{\ensuremath{{M}^{\mathit{ss}}}\xspace}

\newcommand{\Pz}{\ensuremath{P_{z}}\xspace}
\newcommand{\Pzeff}{\ensuremath{P_{\zeff}}\xspace}
\newcommand{\Px}{\ensuremath{P_{x}}\xspace}
\newcommand{\Pxeff}{\ensuremath{P_{\xeff}}\xspace}

\newcommand{\Zs}{\ensuremath{\tilde{Z}}\xspace}
\newcommand{\dZs}{\ensuremath{\Delta\tilde{Z}}\xspace}
\newcommand{\Z}{\ensuremath{Z}\xspace}
\newcommand{\Zss}{\ensuremath{{Z^{\mathit{ss}}}}\xspace}
\newcommand{\Zssref}{\ensuremath{{Z^{\mathit{ss}}(-\dw)}}\xspace}
\newcommand{\Zsslab}{\ensuremath{{Z^{\mathit{ss}}(+\dw)}}\xspace}

\newcommand{\lameff}{\ensuremath{\lambda_{\mathit{eff}}}\xspace}

\newcommand{\li}{\ensuremath{\lambda_{1}}\xspace}

\newcommand{\Reff}{\ensuremath{R_{\mathit{eff}}}\xspace}
\newcommand{\Rho}{\ensuremath{R_{1\rho}}\xspace}
\newcommand{\Rex}{\ensuremath{R_{\mathit{ex}}}\xspace}
\newcommand{\Rexmax}{\ensuremath{R_{\mathit{ex}}^{\mathit{max}}}\xspace}
\newcommand{\Rexlab}{\ensuremath{R_{\mathit{ex}}}(+\dw)\xspace}
\newcommand{\Rexref}{\ensuremath{R_{\mathit{ex}}(-\dw)}\xspace}

\newcommand{\Robs}{\ensuremath{R_{\mathit{obs}}}\xspace}
\newcommand{\mo}{\ensuremath{M_0}\xspace}
\newcommand{\vmo}{\ensuremath{\vec{M}_0}\xspace}
\newcommand{\moa}{\ensuremath{M_{\mathit{0,a}}}\xspace}
\newcommand{\mob}{\ensuremath{M_{\mathit{0,b}}}\xspace}

\newcommand{\tsat}{\ensuremath{t_{\mathit{sat}}}\xspace}

\newcommand{\rf}{\textit{rf}\xspace}
\newcommand{\SL}{SL\xspace}

\newcommand{\CEST}{CEST\xspace}
\newcommand{\BM}{BM\xspace}
\newcommand{\pa}{\textit{a}\xspace}
\newcommand{\pb}{\textit{b}\xspace}
\newcommand{\invivo }{\textit{in vivo}\xspace}
\newcommand{\MTRasym}{\ensuremath{\mathrm{MTR}_{\mathrm{asym}}}\xspace}
\newcommand{\MTR}{\ensuremath{\mathrm{MTR}}\xspace}
\newcommand{\MTRasymss}{\ensuremath{\mathrm{MTR}_{\mathrm{asym}}^{\mathit{ss}}}\xspace}

\usepackage{color}

\newcommand{\add}[1]{\textcolor{black}{#1}}

\usepackage{ulem}
\usepackage{amssymb}
\usepackage{amsthm}
\usepackage{amsmath}

\journal{Journal of Magnetic Resonance}

\begin{document}

\begin{frontmatter}

\title{Exchange--dependent relaxation in the rotating frame for slow and intermediate exchange -- Modeling off-resonant spin-lock and chemical exchange saturation transfer}
\author[label1]{Moritz Zaiss}
\author[label2]{Peter Bachert}
\date{\today}
\address[label1]{Corresponding author, Department of Medical Physics in Radiology, German Cancer Research Center (DKFZ), Im Neuenheimer Feld 280, 69120 Heidelberg, Germany, Phone: +49 6221 422543, Fax: +49 6221 422531, m.zaiss@dkfz.de}
\address[label2]{Department of Medical Physics in Radiology, German Cancer Research Center (DKFZ), Im Neuenheimer Feld 280, 69120 Heidelberg, Germany}

\begin{abstract}
Chemical exchange observed by NMR saturation transfer (CEST) or spin-lock (SL) experiments provide a  MR imaging contrast  by indirect detection of exchanging protons. Determination of relative concentrations  and exchange rates are commonly achieved by  numerical integration of the Bloch-McConnell equations.
We derive an analytical solution of the Bloch-McConnell equations that describes the magnetization of coupled spin populations  under radio frequency irradiation. As CEST and off-resonant SL are equivalent, their steady-state magnetization and the dynamics can be predicted by the same single eigenvalue which is the longitudinal relaxation rate in the rotating frame \Rho. For the case of slowly exchanging systems, e.g. amide protons, the saturation of the small proton pool is affected by transversal relaxation (\R{2b}). It comes out, that \R{2b} is also significant for intermediate exchange, such as amine- or hydroxyl-exchange, if  pools are only partially saturated. We propose a solution for \Rho that includes \R{2b} of the exchanging pool by extending existing approaches and verify it by numerical simulations.
With the appropriate projection factors we obtain an analytical solution for CEST and SL for non-zero \R{2} of the exchanging pool, exchange rates in the range of 1 to $10^4$~Hz, \bi from $0.1$ to $10~\mu$T,  arbitrary chemical-shift differences between the exchanging pools, while considering the dilution by direct water saturation across the entire Z-spectra. This allows  optimization of irradiation parameters and quantification of pH-dependent exchange rates and metabolite concentrations. Additionally, we propose evaluation methods that correct for concomitant direct saturation effects. It is shown that existing theoretical treatments for CEST are special cases of this approach. 
\end{abstract}
\begin{keyword}
spin-lock\sep magnetization transfer\sep Bloch-McConnell equations\sep chemical exchange saturation transfer\sep PARACEST \sep HyperCEST
\end{keyword}

\end{frontmatter}


\section{Introduction}
\label{seq:intro}

The relaxation of an abundant spin population is affected by a rare spin population owing to inter- and intramolecular magnetization transfer processes mediated by scalar or dipolar couplings or chemical exchange \cite{wolff_nmr_1990}. As a consequence, by selective radio frequency (\rf) irradiation of a coupled rare population not only the relaxation dynamics, but also the steady-state magnetization of the abundant population can be manipulated. Due to this preparation, the NMR signal of the abundant population contains additional information on the rare population and its interactions. 
In this context, we analyze two experiments : chemical exchange saturation transfer (\CEST) \cite{zhou_chemical_2006} and off-resonant spin-lock (\SL). 

\CEST and \SL experiments are commonly applied to enhance the NMR sensitivity of protons in diluted metabolites \invivo  \cite{zhou_using_2003,cai_magnetic_2012,jin_magnetic_2012,ling_assessment_2008} yielding an imaging contrast for different pathologies \cite{jia_amide_2011,schmitt_new_2011,zhou_amide_2011,gerigk_7_2012,schmitt_cartilage_2011}. 
The normalized z-magnetization after irradiation at different frequencies, the so-called Z-spectrum, is affected by relaxation and irradiation parameters. In the following, the large pool of water protons is called pool \pa and the pool of \add{dilute} protons pool \pb. 
To obtain a pure contrast that depends only on the exchanging pool \pb, concomitant effects like direct water saturation or partial labeling of the exchanging proton pool must be taken into account in modeling of Z-spectra.
Similarities between CEST and SL have been noticed before \cite{jin_spin-locking_2011,vinogradov_pcest:_2012}. Here we consider the projection factors which are required for  application of static and dynamic solutions derived for \SL to \CEST experiments and vice versa. We demonstrate how the experimental data have to be normalized that the dynamics of \CEST and \SL can be described by one single eigenvalue, namely \Rho, the longitudinal relaxation rate in the rotating frame. A first approximation for \Rho including chemical exchange was published by Trott and Palmer \cite{trott_r1rho_2002}. \add{In the present article, this approach is extended by inclusion of \R{2b}, the transverse relaxation rate of pool \pb.}

An interesting CEST effect is amide proton transfer (APT) of $^1H$ in the backbone of proteins, because quantitative determination of the exchange rate may allow noninvasive pH mapping \cite{sun_imaging_2008}. 
The exchange rate \kb for APT is relatively small (\kb=$28.6\pm 7.4$~Hz \cite{zhou_chemical_2006}) compared to the transversal relaxation rate of the amide proton pool $\R{2b}=1/T_{2b}$. Sun et al. measured $T_{2b}$ of 8.5~ms ( $\R{2b}=90.9~\text{Hz}$) for amine protons of aqueous creatine at \B=9.4~T. For amino protons in ammonium chloride dissolved in agar gel, $T_{2b}=40 \text{ ms } (\R{2b}=25 \text{ Hz}$) was found at \B=3~T \cite{desmond_understanding_2012}. Thus, \R{2b} in tissue may be in the range of or even surpass \kb and must be taken into account for quantification of \kb.
For systems with strong hierarchy in the eigenvalues - as it is the case for diluted spin  populations - we present an approximation for \Rho that includes \R{2b} and provide an analytical solution for \CEST  and \SL experiments valid for exchange rates in the range of \R{2b}. 

\section{Theory}
\label{seq:theo}

\CEST  and \SL experiments for coupled spin systems can be described by classical magnetization vectors \Mv in Euclidean space governed by the Bloch-McConnell (\BM) equations \cite{mcconnell_reaction_1958}. We consider a system of two spin populations: pool \pa (abundant pool) and pool \pb (rare pool) in a static magnetic field $\vec{\B} = (0, 0, \B)$, with \add{forward rate} \kb and thermal equilibrium magnetizations \moa and \mob, respectively. The relative population fraction $\frac{\mob}{\moa}=\fb$  is conserved by the back exchange rate $\ka = \fb\kb$. 

The 2-pool \BM equations are six coupled first-order linear differential equations
 \begin{align}
\label{eqn:BM_diff}
\dot{\vec{M}}=\mathbf{A}\cdot \vec{M} + \vec{C}, \quad 
\A=
\begin{bmatrix}
\mathbf{L}_a-\fb \mathbf{K} & +\mathbf{K}	\\ 
+\fb \mathbf{K}	& \mathbf{L}_b-\mathbf{K}
\end{bmatrix},
\end{align}
where (i = a,b)
\begin{align}
\mathbf{L}_i=
\begin{pmatrix}
-\R{2{i}} 	 	& -\dw_{i}  			&0  	\\ 
 +\dw_{i}  				&-\R{2{i}} 	&-\wi \\
0      				&+\wi 			&-\R{1{i}}
\end{pmatrix}, \quad 
\label{eqn:BM_KL}
\mathbf{K}=
\begin{pmatrix}
\kb &0 &0	\\ 
0 &\kb &0	\\ 
0 &0 &\kb	 
\end{pmatrix}, \\
\label{eqn:BM_C}
 \vec{C}=\begin{pmatrix}
&0, &0, &\R{1a}\moa, &0, &0, &\R{1b}\mob
\end{pmatrix}^\text{T},
\end{align}
given in the rotating frame $(x,y,z)$ defined by \rf irradiation with frequency \wrf. $\dw=\dw_a=\wrf-\wa$  is the frequency offset relative to the Larmor frequency \wa of pool \pa  (for $^1H \; \wa/\B = \gamma = \mathrm{267.5~\frac{rad}{\mu T s}}$). The offset of pool \pb $\dwb=\wrf-\wb=\dw-\db\wa$ is shifted by  \db (chemical shift) relative to the abundant-spin resonance. \add{In contrast to Ref. \cite{trott_r1rho_2002}, we allow different relaxation rates \R{1} and \R{2} for the pools. The assumption of their equality is only valid if $|\R{1a}-\R{1b}|\ll\kb$ or $|\R{2a}-\R{2b}|\ll\kb$ \cite{miloushev_r1_2005}.} Longitudinal relaxation rates $\R{1,a/b} = 1/T_{1,a/b}$ are  in the order of Hz, while transverse relaxation rates $\R{2,a/b} =1/T_{2,a/b}$ are   10-100~Hz. For semisolids \R{2b} can take values up to $10^6$~Hz. The \rf irradiation field $\vec{\bi} = (\bi, 0, 0)$ in the rotating frame, with $\bi \approx \mu T$, induces a precession of the magnetization with frequency $\wi=\gamma\cdot\bi$ around the x-axis in the order of several 100 Hz. The population fraction \fb is assumed to be $<1~\%$, hence \ka is   0.01 to 10~Hz. 

\subsection{\add{Solution of the Bloch-McConnell equations for asymmetric populations}}
\begin{figure}[H]
\includegraphics{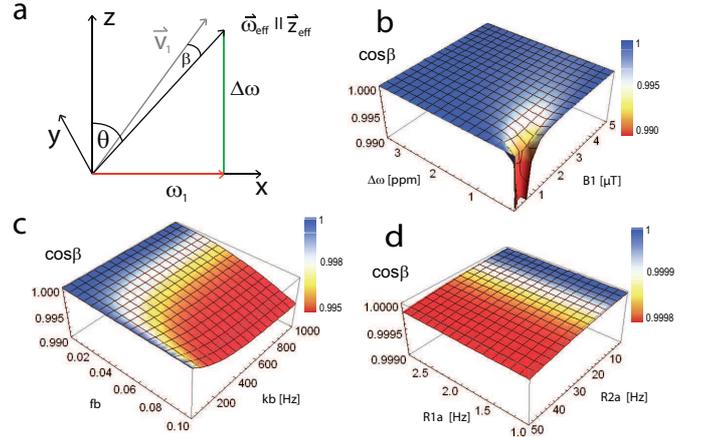}
\caption{(a) Geometry of the vectors in the rotating frame. (b-d) \add{Cosine of the angle $\beta$ between} the eigenvector of the smallest eigenvalue and \vweff. (b) In the far off-resonant case both vectors are parallel. Near resonance \wi has to be strong to keep them parallel. (c) The assumption of collinearity is still valid if pool \pb with relative concentration $\fb < 10\%$ is coupled to the water pool. (d) Large differences in \R{2a} and \R{1a} lead to an increasing angle between the vectors, but even for $\R{2a}\approx 50Hz$ and $\R{1a}\approx 1$Hz both vectors are still collinear in good approximation.
The eigenvector and the effective field vector are collinear if \weff is large compared to $(\R{2a}-\R{1a})$ \add{(\ref{seq:App_EV})} and $\fb < 10\%$ -- both is fulfilled for CEST experiments since metabolite concentrations are small and frequency offsets of interest are mostly larger than several 100 rad/s. }
\label{fig:EVweff}
\end{figure}

The BM equations \eqref{eqn:BM_diff} are solved in the eigenspace of the matrix \A leading to the general solution for the magnetization	
 \begin{align}
\label{eqn:BM_generalsol}
\vec{M}(t)=\sum_{n=1}^{6}e^{\lambda_n t}\vec{v}_n+\vmss ,
\end{align}
where $\lambda_n $ is the nth eigenvalue with the corresponding eigenvector $\vec{v}_n$  of matrix \A and \vmss    is the stationary solution. Two eigenvalues are real and four are complex \cite{trott_r1rho_2002}. They describe precession and, since all real parts of the eigenvalues are negative, the decay of the magnetization towards the stationary state in each pool.
As shown before  \cite{trott_theoretical_2004}, if \dw  or  \wi are large compared to the relaxation rates \R{1} and \R{2} and exchange rate \kb, the eigensystem of pool \pa is mainly unaffected. One \add{eigenvector} $\vec{v}_1$ is closely aligned with the effective field $\vweff=(\wi,0,\dw)$ which defines the longitudinal direction (\zeff) in the effective frame (\xeff,\yeff,\zeff) and is tilted around the y-axis by the angle $\theta = \tan^{-1}(\frac{\wi}{\dw})$ off the z-axis (Fig. \ref{fig:EVweff}a). \add{Mathematical derivation (\ref{seq:App_EV}) as well as} numerical evaluations (Fig.\ref{fig:EVweff}b-d) demonstrate that $\vec{v}_1$ and \vweff are collinear in good approximation if $(\R{2a}-\R{1a})$ is much smaller than \weff. 

The collinearity of the corresponding eigenvector and the effective field is the principal reason why off-resonant \SL and \CEST exhibit the same dynamics. For an appropriate analysis of a saturation experiment it is mandatory to identify the initial projections on the eigenvectors and the measured components.
\vB and \vmo are parallel to the z-axis, the preparation is a projection of the longitudinal magnetization along~z onto the effective frame
 \begin{align}
\label{eqn:hintrafoz}
M_{\zeff}(t=0)&=\cos{\theta} \cdot M_z(t=0)=\Pzeff \cdot \mo ,   \\
\label{eqn:hintrafoxy}
M_{\xeff}(t=0)&=\sin{\theta} \cdot \mo ;\quad M_{\yeff}(t=0)=0 .
\end{align} 
The transversal components induce an oscillation decaying with $T_{2\rho}$ \add{\cite{moran_near-resonance_1995}} which can be neglected in the case of small $\theta$, by averaging over a complete cycle of \weff, or by measuring after a delay of $~5\cdot T_{2\rho}$. This simplification leads to the relation for the back projection, via \Pz, from \zeff to~z
 \begin{align}
\label{eqn:backtrafoz}
M_{z}(t)=\cos{\theta} \cdot M_{\zeff}(t)=\Pz \cdot  M_{\zeff}(t).
\end{align} 
Since we identified the effective frame as the eigenspace of the magnetization, Eq.~\eqref{eqn:BM_generalsol} can be written as an exponential decay law with the eigenvalue \li associated with the \zeff~direction. Let the normalized magnetization be $\Z=\frac{M_{\mathit{z,a}}}{\moa}$ and, for the stationary solution, $\Zss=\frac{\mss_{\mathit{z,a}}}{\moa}$. Then Eq.~\eqref{eqn:BM_generalsol}, taken for the \zeff direction, yields the dynamic solution for the z-magnetization 
 \begin{align}
\label{eqn:Z_full_solution_apex}
\Z(\dw,\wi,t)=(\Pz\Pzeff-\Zss)\cdot e^{\li\cdot t}+\Zss
\end{align}

\begin{figure}[H]
\begin{center}
\includegraphics{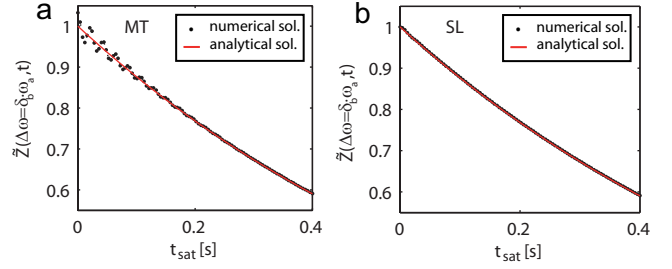}
\end{center}
\caption{The full numerical Bloch-McConnell solution (dots) with the proposed normalization (Eq.~\eqref{eqn:Zs}) demonstrates the equivalence of two experiments: chemical exchange saturation transfer (\CEST) without preparation pulses (a); spin-lock (\SL) with preparation and measurement in the effective frame (b). \Zs of \CEST undergoes oscillations because of residual transversal magnetization in the effective frame. \SL shows no oscillations since the transversal magnetization in the effective frame is zero [\dZs= 0, see text] (Eq.~\eqref{eqn:dZs}). Both, \SL and \CEST, show the same monoexponential decay of the z-magnetization with \li (Eq.~\eqref{eqn:SOL_l1_leff_lstrich}) (solid red). For full BM simulations \cite{woessner_numerical_2005} parameters were taken from the amide proton system \cite{zhou_using_2003} in brain white matter  \cite{stanisz_t_2005} at \B = 3 T: If not varied, $\R{2a} = 14.5$~Hz, $\R{1a}= \R{1b}  = 0.954$~Hz, $\R{2b} = 66.6$~Hz, $\fb = 1$~\%, $\kb = 25$~Hz, $\db = 3.5$~ppm, $\bi= 1~\mathrm{\mu T}$, $\tsat=1$~s.}
\label{fig:1}
\end{figure}

Without preparation pulses $\Pz = \Pzeff  \approx \cos{\theta}$ (\CEST experiment). 
If a preparation pulse with flip angle $\theta$ is applied before and after cw irradiation  the projection factors are $\Pz = \Pzeff \approx 1$ (\SL experiment), hence oscillations are suppressed (Fig.~\ref{fig:1}), but still persist since \zeff is not perfectly collinear with the eigenvector. 
Transformation of Eq.~\eqref{eqn:BM_diff} into the effective frame and setting $\frac{\text{d}}{\text{d}t}\Mv = 0$ yields the steady-state solution \add{(\ref{seq:App_EV})}
 \begin{align}
\label{eqn:Zss}
\Zss(\dw,\wi)=-\frac{\Pz\cdot R_{1a}\cdot\cos{\theta}}{\lambda_1}.
\end{align}
It is important to note that in the case where the steady-state is non-zero, it is locked along the corresponding eigenvector. Equations \eqref{eqn:Z_full_solution_apex} and \eqref{eqn:Zss} agree with the full solution previously found for \SL by Jin et al. \cite{jin_magnetic_2012} but extend it for \CEST.

\add{To obtain a pure dynamic quantity independent of the steady-state we rearrange Eq.\eqref{eqn:Z_full_solution_apex} \add{and define}}
 \begin{align}
\label{eqn:Zs}
\Zs(\dw,\wi,t)\equiv\frac{Z-\Zss}{\Pz\Pzeff-\Zss}=e^{\li\cdot t}.
\end{align}
\add{Eqs. \eqref{eqn:Zss} and \eqref{eqn:Zs} are the central formulas in this article.}

In fact, the description of \SL and \CEST experiments differs in the projection factors \Pz and \Pzeff. The intuitive solution $Z_{CEST}=\cos{\theta}\cdot Z_{SL}$ is valid for the steady-state, but not for the transient-state. If the initial magnetization $M_i$ is not fully relaxed and flipped before the saturation pulse by an angle $\beta$, \Pzeff changes to $\cos(\theta-\beta)\cdot{M_i/\mo}$.

After understanding of the transition between the two experiments we will now solve the dynamics of CEST and SL experiments by finding the corresponding eigenvalue and verify it numerically.
As already demonstrated for the \SL experiment  \cite{trott_r1rho_2002}, the eigenvalue, which corresponds to the eigenvector along the \zeff-axis, is the smallest eigenvalue in modulus of the system.
Assuming that all eigenvalues of an arbitrary full-rank matrix \A are much larger in modulus than the smallest eigenvalue, i.e. $|\lambda_1|\ll|\lambda_{2...n}|$, we obtain (see \ref{seq:App})  
\begin{align}
\label{eqn:c0c1}
\lambda_1\approx-\frac{c_0}{c_1} , 
\end{align}
where $c_0$ and $c_1$ are the coefficients of the constant and the linear term of the normalized characteristic polynomial, respectively.
We derive the full solution for the smallest eigenvalue by employing the solution of the unperturbed system $(\fb=0)$. The solution is $\lameff=-\Reff$ with the decay rate in the effective frame \Reff which was shown to be approximately \cite{trott_theoretical_2004}
 \begin{align}
\label{eqn:reff}
-\Reff = \R{1a}\cos^2{\theta}+\R{2a}\sin^2{\theta}.
\end{align}

With this eigenvalue of the unperturbed system we can rescale the system by 
\begin{align}
\label{eqn:Astrich_rescale}
\As=\A-\mathbf{I}\cdot\lameff
\end{align}

\add{thus shifting the smallest eigenvalue by \Reff. The smallest eigenvalue of \As, still contains terms of \R{1a} and \R{2a}, but represents the exchange-induced perturbation of \Rho.}

\begin{figure}[H]
\begin{center}
\includegraphics{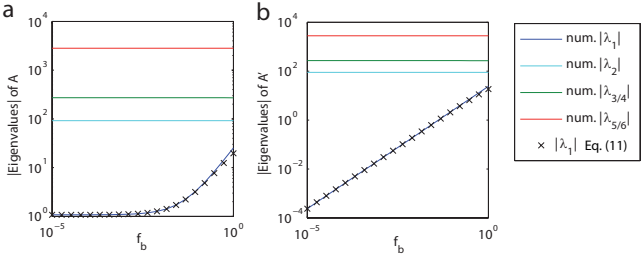}
\end{center}
\caption{Hierarchy of numerically calculated BM eigenvalues (lines) of the standard system (see caption of Fig. \ref{fig:1}). (b) The rescaled matrix \As (eq. \eqref{eqn:Astrich_rescale}) has a much stronger hierarchy in the eigenvalues than matrix \A (a). This improves the approximation  of the smallest eigenvalue (x, Eq. \eqref{eqn:c0c1}).  }
\label{fig:EW}
\end{figure}

 The result is a strong hierarchy (Fig. \ref{fig:EW}) in the eigenvalues of \As if the coupling is small ($\fb\ll 1$). Now Eq.~\eqref{eqn:c0c1} can be employed to calculate the eigenvalue $\lambda_1'$ of the matrix \As to obtain the full solution:

 \begin{align}
\label{eqn:SOL_l1_leff_lstrich}
\lambda_{1}=\lameff+\lambda_{1}' .
\end{align}
Here $\lambda_{1}' =-c_0'/c_1'$ is the ratio of the coefficients of the characteristic polynomial of the matrix \As. This analytical procedure gives us a very good approximation of the dynamics of the \BM system.

For further simplification we assume that relaxation of pool \pa is well described by \Reff and the perturbation is dominated by the exchange and relaxation of pool \pb. We call the exchange-dependent relaxation rate $\Rex=-\lambda_1'$. The eigenvalue \li is associated with \zeff \add{and} is therefore an approximation of the relaxation rate in the rotating frame $\Rho\approx-\li$  given by Eqs. \eqref{eqn:reff} and \eqref{eqn:SOL_l1_leff_lstrich}
 \begin{align}
\label{eqn:SOL_Rho_reff_rex}
\Rho(\dw) = \Reff(\dw) +\Rex(\dw) .
\end{align}
To derive a useful approximation of \Rex, we neglect all relaxation terms of pool \pa in matrix \As. Furthermore, we assume that \R{1b} is much smaller than \R{2b} and \kb and therefore \R{1b} can be neglected in \As. In contrast to Trott and Palmer \cite{trott_r1rho_2002}, we do not neglect \R{2b}, but \ka. By this means, the obtained eigenvalue approximation by using Eq.~\eqref{eqn:c0c1} is linearized in the small parameter \fb giving
 \begin{align}
\label{eqn:SOL_Lorentz}
\Rex(\dwb)=  \frac{\Rexmax\frac{\Gamma^2}{4}}{\frac{\Gamma^2}{4}+\dwb^2} 
\end{align}
with maximum value
 \begin{align}
\label{eqn:SOL_Rexmaxfull}
\Rexmax= \fb \kb\sin^2{\theta}   
\frac{ (\wb-\wa)^2+ \frac{\R{2b}}{\kb}(\wi^2+\dw^2)+ \R{2b}(\kb+\R{2b}) }{\frac{\Gamma^2}{4}} 
\end{align}
and full width at half maximum (FWHM)  
 \begin{align}
\label{eqn:SOL_RexGamma}
\Gamma=2\sqrt{\frac{\kb+\R{2b}}{\kb}\wi^2+(\kb+\R{2b})^2  } .
\end{align}
For large $|\wb-\wa|$ 
 \begin{align}
 \label{eqn:SOL_RexmaxLS}
\Rexmax\approx\fb \kb\cdot \frac{\wi^2}{\wi^2+\kb(\kb+\R{2b})} .
\end{align}
The \wi-dependent factor yields the amount of labeling of pool \pb. Hence, we call this factor  labeling efficiency, refering to \cite{sun_correction_2007}:
\begin{align}
 \label{eqn:alpha}
\alpha=\frac{\wi^2}{\wi^2+\kb(\kb+\R{2b})} .
\end{align}

For strong \bi and small \R{2b} and \kb, $\alpha$ is approximately one and we obtain the \textit{full-saturation} limit
 \begin{align}
\label{eqn:SOL_Rexmax_ka}
\Rexmax\approx\fb \kb=\ka .
\end{align}

\section{Results}
\label{seq:res}
We obtained numerical values for the eigenvalues computed by means of the full numerical BM matrix solution  \cite{woessner_numerical_2005} and compared them to the proposed approximations via
 \begin{align}
\label{eqn:numanaexp_rex}
\Rex =-|\lambda_{1,\text{numerical}}|-\Reff .
\end{align}
To verify equations (\ref{eqn:Zs},\ref{eqn:Zss},\ref{eqn:SOL_l1_leff_lstrich},\ref{eqn:SOL_Rexmaxfull}) the dynamics of the magnetization vectors of the exchanging spin pools were simulated. The decay rate \Rex is obtained from \Zs (Eq. \eqref{eqn:Zs} ) and \Reff via
 \begin{align}
\label{eqn:exp_rex}
\Rex =-\frac{\log(\Zs)}{\tsat}-\Reff .
\end{align}
The simulation parameters for the abundant pool were chosen according to published data for brain white matter  \cite{stanisz_t_2005} including a rare pool attributed to amide protons  \cite{zhou_chemical_2006}.

\begin{figure}[H]
\begin{center}
\includegraphics{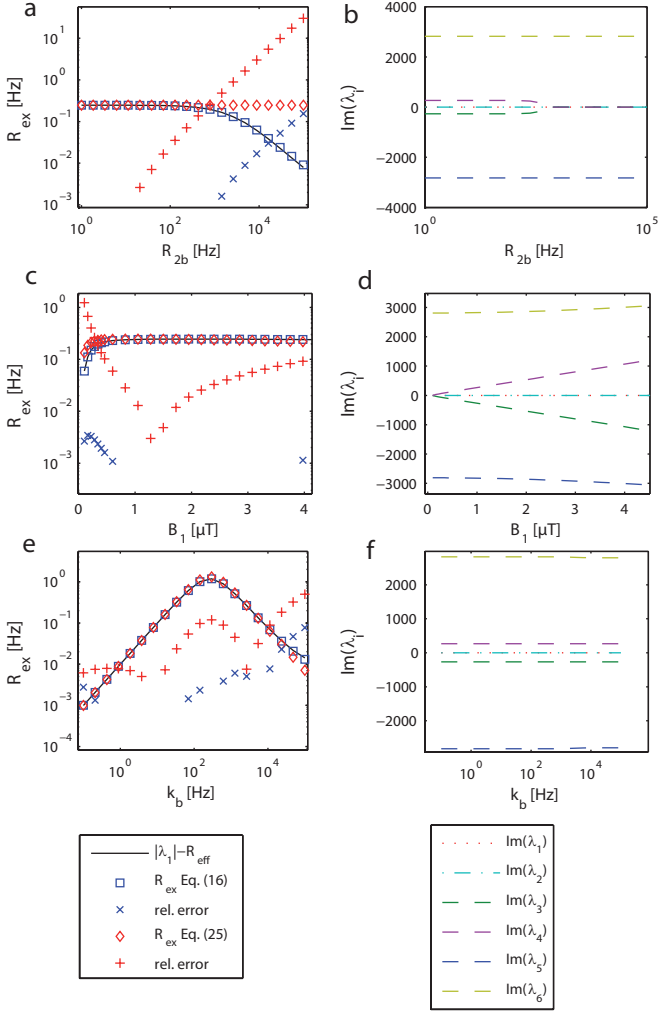}
\end{center}
\caption{(a,c,e) \Rex on-resonant on pool \pb from smallest eigenvalue in modulus (Eq. \eqref{eqn:numanaexp_rex})  , calculated numerically (line) and analytically by the approximations of Eq. \eqref{eqn:SOL_Lorentz} (squares)  and the asymmetric population limit of Trott and Palmer \cite{trott_r1rho_2002} (Eq. \eqref{eqn:SOL_Rho_Trott}, diamonds). x and + mark the relative error $(1 - (\Rex^{ana}/\Rex^{num}))$ when it is larger than 0.1 \%. (a) For small \R{2b}, both solutions for \Rex agree with the numerical value; if \R{2b} is larger than \kb the proposed solution still matches the numerical value. The extension by \R{2b} is important if the CEST pool is not fully saturated, which is the case for small \bi (c) or large \kb (e). But also for large \bi the solution, that includes \R{2b} fits the numerical value with higher accuracy.  (b,d,f) Imaginary parts of the numerical eigenvalues. \bi and \R{2b} ranges where Eq. \eqref{eqn:SOL_Lorentz} (squares) shows deviations from the numerical solution correlate with ranges where the imaginary part becomes small or even zero. In this case, the assumption of a strong hierarchy in the eigenvalues is not valid anymore.}
\label{fig:EW_trott_zaiss}
\end{figure}

The proposed approximation of \Rex by Eq.\eqref{eqn:SOL_Lorentz} was compared to the asymmetric population solution of Ref.\cite{trott_r1rho_2002} (Fig. \ref{fig:EW_trott_zaiss}). If \R{2b} is non-zero, \Rex proposed  by Eq.\eqref{eqn:SOL_Lorentz} matches the numerical value better than the \Rex \add{given in Ref.\cite{trott_r1rho_2002}} \add{(see Eq. \eqref{eqn:SOL_Rho_Trott} below)}.
Especially the dependence of \Rex on \bi (Fig. \ref{fig:EW_trott_zaiss}c) changes by taking \R{2b} into account.

\begin{figure}[H]
\begin{center}
\includegraphics{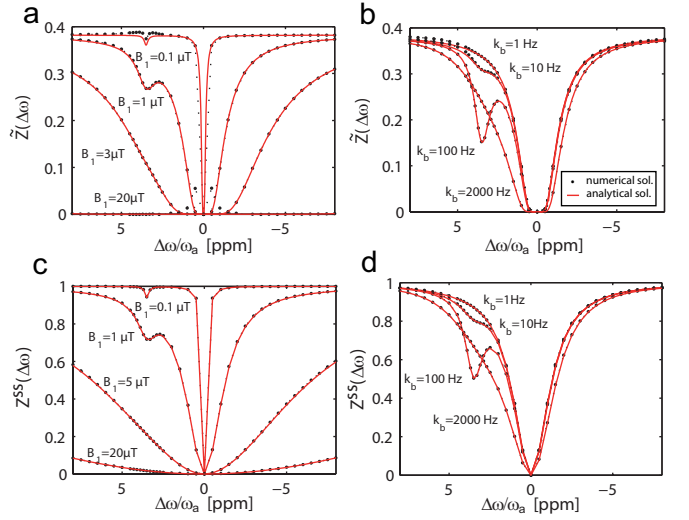}
\end{center}
\caption{Numerical simulations (dots) of an \CEST experiment evaluated for dynamic \Zs-spectra (a,\,b) and steady-state Z-spectra (c,\,d) are in agreement with Eqs.~\eqref{eqn:Zs} and \eqref{eqn:Zss} (solid red), respectively. Plots demonstrate high correlation for different $\bi (0.1 - 20~\mathrm{\mu T}; a, c)$ and $\kb(1 - 2000~\mathrm{Hz}; b, d)$. Deviations near resonance of pool \pa for large \bi (a) are caused by oscillations of the magnetization in the transverse plane of the effective frame. In the \SL experiment these oscillations are suppressed (Fig.~\ref{fig:1}).}
\label{fig:ZZs_spectra}
\end{figure}
For an \CEST experiment, the normalized numerical solution agrees with the theory of dynamic \Zs-spectra (Fig.~\ref{fig:ZZs_spectra}a,\,b) and steady-state Z-spectra (Fig.~\ref{fig:ZZs_spectra}c,\,d) for different values of \bi and \kb. The competing direct and exchange-dependent saturation -- a central problem in proton CEST \cite{sun_imaging_2008,sun_correction_2007,zaiss_quantitative_2011} -- is modeled correctly. Deviations in Fig.~\ref{fig:ZZs_spectra}a for strong \bi and $\dw \rightarrow 0$ result from transversal magnetization in the effective frame which was neglected before. By projection on the transverse plane of the effective frame using Eqs.~\eqref{eqn:hintrafoxy} we obtained the resulting deviation of \Zs   
 \begin{align}
\label{eqn:dZs}
\dZs(\dw,\wi,t)=\frac{\Px\Pxeff}{\Pz\Pzeff-Z_{ss}}\cdot Re(e^{\lambda_2\cdot t}) ,
\end{align}
with projections \Pxeff and \Px into the transverse plane of the effective frame and back. For MT $\Px=\Pxeff=\sin{\theta}$.  Real and imaginary parts of the complex eigenvalue $\lambda_2$ are given by  $-\R{2\rho}\approx-\frac{1}{2}(\R{2a}+\R{1a}\sin^2{\theta}+\R{2a}\cos^2{\theta})$  \cite{moran_near-resonance_1995} and \weff, respectively. The implicit neglect of \dZs in Eq.~\eqref{eqn:Zs} is justified if $\tsat\gg T_{2\rho}$ or \Px and \Pxeff are small. This can be realized either by \SL preparation or by $\wi\ll\dw$.
The on-resonant case of \CEST $(\theta = 90^\circ) $ is not defined, because \Zss in Eq.~\eqref{eqn:Zss} and thus the denominators in Eqs.~\eqref{eqn:Zs} and \eqref{eqn:dZs} vanish. Then the z-axis lies in the transverse plane of the effective frame and  Z is described by $\moa\cdot Re(e^{(-(\R{2\rho}+i\wi) t)}) $.
Therefore, near resonance \SL is preferable to \CEST; it also yields in general a higher SNR (given by the projection factors \Pz, \Pzeff). Regarding the experimental realization, \CEST is simpler than \SL, because  \dw and \wi and thus $\theta$  can be corrected effectively after the measurement by \B and \bi field mapping \cite{sun_correction_2007,kim_water_2009}. In contrast, \SL requires knowledge of \bi and \B during the scan for proper preparation or techniques that are insensitive to field inhomogeneities such as adiabatic pulses  \cite{mangia_rotating_2009,michaeli_transverse_2004}.
\begin{figure}[H]
\includegraphics{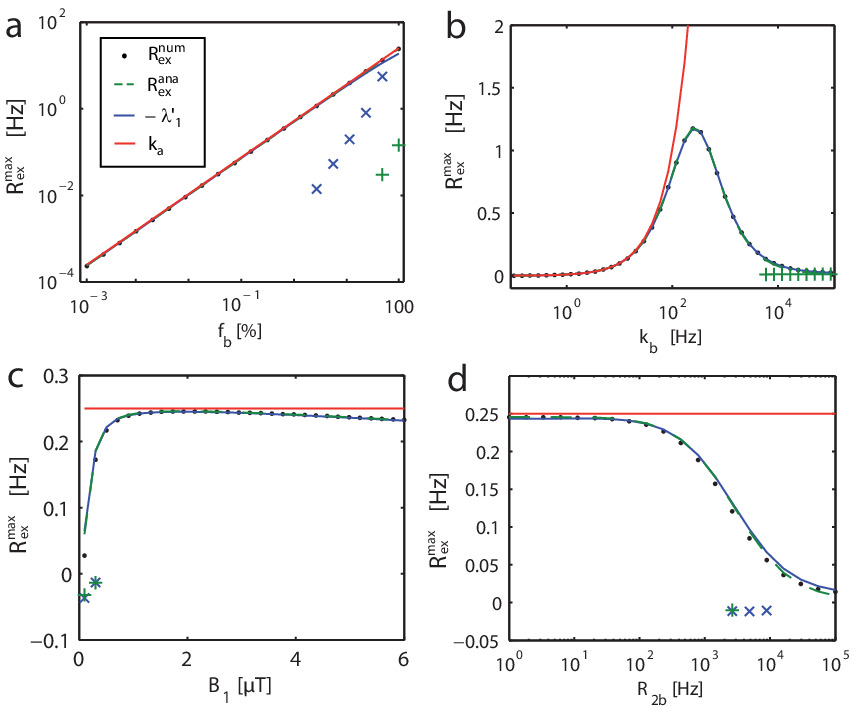}
\caption{Numerical \Rexmax (dots), employing Eq.~\eqref{eqn:exp_rex}, fit to $-\lambda_1'$ of Eq.~\eqref{eqn:SOL_l1_leff_lstrich} (solid blue) and to \Rexmax of Eq.~\eqref{eqn:SOL_Rexmaxfull} (dashed green) as a function of \fb, \kb, \bi, and \R{2b}. x and + mark the relative error $(1 - (\Rex^{ana}/\Rex^{num}))$ when it is larger than 1 \%. (a) As expected, the eigenvalue approximation is insufficient for $\fb  > 5\% $. (b) Contrary to the approximation of Eq.\eqref{eqn:SOL_Rexmax_ka} (solid red) the full solution follows the decrease of \Rexmax with large \kb. The decrease of deviations for small \bi (c) and high \R{2b} (d) may be caused by overdamping in pool b, i.e., eigenvalues become real which reduces the required hierarchy in the set of the eigenvalues. However, the deviation in \Rex is too small (Fig. \ref{fig:EW_trott_zaiss}) to explain this deviation leading to the conclusion that other eigenvectors are contributing to the relaxation. (d) Inclusion of \R{2b} is relevant for $\R{2b} > 100$ Hz. } 
\label{fig:Rex}
\end{figure}
The values of the rate \Rexmax obtained by simulations fit well to the full (Eq.~\eqref{eqn:SOL_l1_leff_lstrich}) and approximate (Eq.~\eqref{eqn:SOL_Rexmaxfull}) solution for the observed parameters (Fig.~\ref{fig:Rex}). Deviations of simulation and analytical solution were smaller than 1\% for rates varied in the ranges: $\R{1b}=0.1-10$ Hz, $\R{1a}=0.1-10$ Hz and $\R{2a}=2-100$ Hz (data not shown).
 
\section{Discussion}
\subsection{General solution}
We showed that our formalism, established by Eqs.~\eqref{eqn:Zs},\eqref{eqn:Zss} and \eqref{eqn:Z_full_solution_apex} together with the eigenvalue approximation of Eq.~\eqref{eqn:SOL_l1_leff_lstrich}, is a general solution for CEST experiments. This now allows us to discuss from a general point of view the techniques and theories proposed in the field of chemical exchange saturation transfer. For the \SL solution this was already accomplished by Jin et al. \cite{jin_spin-locking_2011,jin_magnetic_2012}.

The proposed eigenvalue approximation assumes the case of asymmetric populations. This restricts its application to systems where the water proton pool is much larger than the exchanging pools -- which is the case for CEST experiments. 
There are many analytical approaches for the smallest eigenvalue (\Rho) of the BM matrix  besides our approach. They use pertubation theory \cite{trott_theoretical_2004}, the stochastic Liouville equation \cite{abergel_markov_2005},  an average magnetization approach \cite{trott_average-magnetization_2003}, and the polynomial root finding algorithm of Laguerre \cite{miloushev_r1_2005}. The latter is even valid in the case of symmetric populations.
However, all these treatments  neglect the transverse relaxation of the exchanging pool. Since in CEST experiments the exchange rates are often quite small (e.g., $\kb\approx 28 Hz$ for APT), \R{2b} cannot be neglected against \kb. 
We chose therefore a simple approach which is suitable for the condition of asymmetric populations  and took \R{2b} into account.
Our approach to find the eigenvalue including \R{2b} is similar to that of  Trott and Palmer \cite{trott_r1rho_2002}.
However, different \R{1} and \R{2} were allowed for the involved pools. In addition, an alternative justification of the relation $\lambda_1=-\frac{c_0}{c_1}$ was obtained, which uses the intrinsic hierarchy of the eigenvalues (\ref{seq:App})  instead of linearization of the characteristic polynomial.
By this means, it turned out that a strong hierarchy of the eigenvalues is necessary for the approximation. The  hierarchy was increased by rescaling the system by the unperturbed eigenvalue \Reff (Fig.\ref{fig:EW}). Thus  the accuracy of the approximation was improved. As the parameter \R{2b} was included  and equations were linearized directly in the small parameter \ka, a formula was obtained (Eq.~\eqref{eqn:SOL_Rho_reff_rex}) that differs from   the asymmetric population limit of Ref. \cite{trott_r1rho_2002}  reading 
 \begin{align}
\label{eqn:SOL_Rho_Trott}
\Rho=\Reff+ \underbrace{\sin^2{\theta}  \frac{(\wb-\wa)^2\frac{\ka\kb}{\ka+\kb}}{\dwb^2+\wi^2+(\ka+\kb)^2}}_{\Rex} .
\end{align}
Equality is reached if \R{2b} is neglected in our approximation and if Eq. \eqref{eqn:SOL_Rho_Trott} is linearized in \ka.
With our extension simulated CEST Z-spectra could be predicted well in a broad range of parameters. Moreover, it turned out that \R{2b} is important if it is in the range of \kb (Fig. \ref{fig:Rex}d). Inclusion of  \R{2b} also allows to model macromolecular magnetization transfer effects with  large \R{2b} values (Fig. \ref{fig:IOPA}c).

\begin{figure}[H]
\includegraphics{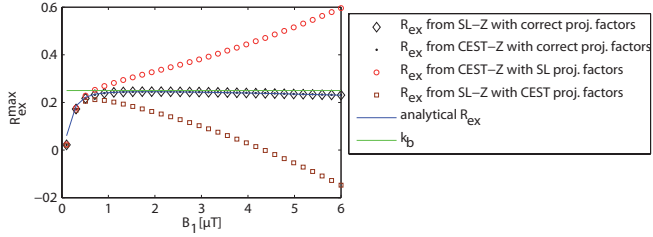}
\caption{The same plot of \Rexmax (Eq. \eqref{eqn:exp_rex}) as in Fig.~\ref{fig:Rex}c, but now for CEST (dots) and SL (diamonds) employing the corresponding projection factors in Eq. \eqref{eqn:Zs} ($\Pz=\Pzeff=1$ for SL and $\Pz=\Pzeff=\cos\theta$ for CEST). Additionally,  the result of an evaluation is shown employing the projection factors of SL for a CEST experiment (circles) and employing the  projection factors of CEST for a SL experiment (squares). Only with the correct projection factors both experiments are described by the same theory and yield \Rex (solid blue). }
\label{fig:wrongP}
\end{figure}

Our solution agrees for \SL with the existing treatment \cite{jin_spin-locking_2011}, but only with the correct projection factors \SL and \CEST can be described by the same theory. This is contrary to the conclusion of Jin et al. \cite{jin_spin-locking_2011} that \SL theory can be used directly to describe CEST experiments. The deviation is not large for small $\theta$, but for $\wi\approx\dw$ the projection factors are crucial as shown in Fig. \ref{fig:wrongP}.

With the correct projections the transition to \CEST is straightforward and provides a much broader range of validity than previous models developed for \CEST which are either appropriate only for small \bi \cite{zaiss_quantitative_2011} or large \bi \cite{baguet_off-resonance_1997} or only \add{for the case of on-resonant irradiation  of pool \pb} \cite{sun_correction_2007,sun_imaging_2008}. The proposed theory (Eq. \eqref{eqn:Z_full_solution_apex}) gives a model for full Z-spectra for transient and steady-state \CEST experiments which enables analytical rather than numerical fitting of experimental data.

\subsection{Extension to other systems}
\begin{figure}[H]
\includegraphics{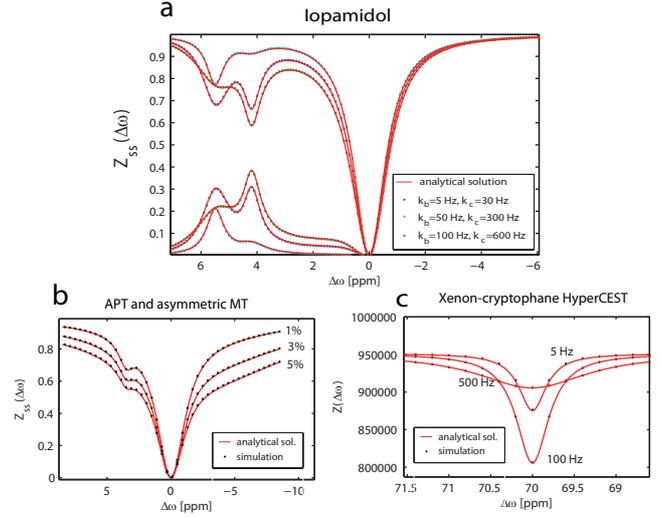}
\caption{Three applications of the proposed theory:(a) The system of iopamidol with corresponding \MTRasym evaluation. (b) The system of APT with an asymmetric macromolecular MT pool \add{(for concentration fractions \fm=1\%, 3\%, 5\%)}. (c) The system of exchanging  hyperpolarized xenon soluted or encapsulated in cryptophane cages, a biosensor method called HyperCEST.}
\label{fig:IOPA}
\end{figure}

As verified for \SL \cite{trott_theoretical_2004}, the theory can be extended to $n$-site exchanging systems. By simply superimposing the exchange-dependent relaxation rates of several pools one obtains the Z-spectra for a multi-pool system. We applied this to the contrast agent iopamidol in water, which has two exchanging amide proton groups \cite{longo_iopamidol_2011}, considering a three pool system: water, amide proton B at 4.2 ppm and amide proton C at 5.5 ppm. Assuming for the exchange rates $k_c=6\cdot\kb$, the superposition of \Reff and the two corresponding \Rex  yields the Z-spectrum of the iopamidol system (Fig. \ref{fig:IOPA}a). A three-pool system relevant for \textit{in vivo} CEST studies includes water protons, amide protons and a macromolecular proton pool. Modeling the macromolecular pool by $\Rex^{m}$(\R{2m} = 5000 Hz, \km = 40 Hz) with an offset of -2.6 ppm and again superimposing it with $\Rex^{amide}$ we are able to model analytically Z-spectra of APT with an underlying symmetric and asymmetric MT effect up to 5\% relative concentration \fm (Fig. \ref{fig:IOPA}b). Hence, the model is able to describe the \textit{in vivo} situation of several CEST pools and underlying MT competing with direct water saturation. Using the superimposed \Rex including $\Rex^{m}$ and $\Rex^{amide}$ and fitting the obtained Z-spectra $\Rex^{amide}$ can be isolated. For macromolecular MT the extension of \Rex by \R{2b} is crucial, since \R{2b} can be as large as $\approx 10^5$ Hz. The implicitly assumed Lorentzian lineshape of the macromolecular pool is only valid around the water proton resonance, for large offsets a super-Lorentzian lineshape must be included in $\Rex^{m}$ \cite{stanisz_t_2005}.

Hyperpolarized xenon spin ensembles exchanging between the dissolved phase and cryptophane cages (HyperCEST experiment, \cite{schroeder_molecular_2006}) can also be described by Eq.~\eqref{eqn:Zs}. Since the initial hyperpolarized magnetization $M_i$ is in the order up to $10^5...10^6\mo$, the steady-state can be neglected for depolarization. This yields $\Z\approx\Zs=M_i \cdot e^{-\Rho\tsat}$ in agreement with the result in Ref. \cite{zaiss_analytical_2012}. Figure ~\ref{fig:IOPA}c shows the simulated Z-spectrum around the cage peak in the HyperCEST experiment of a Xe-cryptophane system for different \kb.

Pulsed irradiation, employed for saturation in SAR limited clinical scanners \cite{schmitt_optimization_2011}, was shown to have similar effects on \MTRasym as cw irradiation with effective \bi \cite{zu_optimizing_2011,sun_simulation_2011}. The presented solution for CEST Z-spectra can therefore be used for optimization of pulsed saturation transfer experiments. 

\subsection{Proton transfer ratio}
For a \CEST experiment the parameters of particular interest are the exchange rate \kb of the metabolite proton pool  and the relative concentration \fb. The former is often pH catalyzed and permits pH-weighted imaging; the latter allows molecular imaging with enhanced sensitivity. The ultimate method must allow -- with high spectral selectivity -- the generation of \kb and \fb maps separately and for different exchanging groups. Unfortunately, both parameters occur in the water pool BM equations as product, i.e. the back-exchange rate $\ka=\kb\cdot\fb$. There are some approaches which are able to separate \kb and \fb for specific cases like rotation transfer of amid protons \cite{zu_multi-angle_2012} or the method of Dixon et al. \cite{dixon_concentration-independent_2010} applicable to PARACEST agents. \add{CEST experiments are commonly evaluated to yield the proton transfer ratio PTR. PTR is an ideal parameter in the sense that it reflects the decrease of the water pool signal owing to exchange from a labeled exchanging pool only, thus neglecting any direct saturation.}

In the following, we assume one CEST pool resonance on the positive \dw axis.

Employing Eq. \eqref{eqn:Zss} with the limit $\theta \rightarrow 0$ we obtain for PTR in steady-state:
\begin{align}
\label{eqn:QUREXtheo_PTR}
\text{PTR}=1-\Zss(\dw) \approx \frac{\Rexlab}{\R{1a}+\Rexlab}
\end{align}

which yields the maximal value $\frac{\ka}{\R{1a}+\ka}$ \cite{zhou_chemical_2006} in the full-saturation limit ($\Rex\approx\ka$ ). Eq. \eqref{eqn:QUREXtheo_PTR} is consistent with PTR including the labeling efficiency $\alpha$ introduced in Ref. \cite{sun_correction_2007}.

\subsection{Z-spectra evaluation - \MTR and \MTRasym}

Methods using asymmetry implicitly assume that the full width at half maximum of \Rex(\dwb) is narrow compared to the chemical shift of the corresponding pool. This means that \Rex(\dwb) can be neglected for the reference scan $\Z(-\dw)$ what is only true in the slow-exchange limit \cite{zhou_chemical_2006}. This limit can be defined more generally by the width of \Rex(\dwb) (Eq. \eqref{eqn:SOL_RexGamma}):
\begin{align}
\label{eqn:slow-ex-limit}
\Gamma=2\sqrt{\frac{\kb+\R{2b}}{\kb}\cdot\wi^2+(\kb+\R{2b})^2} \ll |\dwb-\dw_a|
\end{align}
This new limit depends on \bi which affects the ability to distinguish different peaks in the Z-spectrum (Fig. \ref{fig:ZZs_spectra}c). The limit is therefore a useful parameter for exchange-regime characterization in saturation spectroscopy.

For \CEST the common evaluation parameters are the magnetization transfer rate $\MTR(\dw) = 1-Z(\dw)$ and the asymmetry of the Z-spectrum $\MTRasym(\dw) = Z(-\dw) - Z(+\dw)$. \MTRasym is generally employed to estimate PTR. 
Using Eq.~\eqref{eqn:Zss} together with Eq. \eqref{eqn:SOL_Rho_reff_rex} we obtain for steady-state Z-spectrum asymmetry
\begin{align}
\label{eqn:asymSS}
\begin{split}
\MTRasymss(\dw) &= \Zss(-\dw)-\Zss(+\dw) \\
&= \frac{(\Rexlab-\Rexref)\cdot\R{1a}\Pz\cos{\theta}}{(\Reff+\Rexref)(\Reff+\Rexlab) }.
\end{split}
\end{align}
The comparison shows that \MTRasymss yields PTR of Eq. \eqref{eqn:QUREXtheo_PTR} only if $\theta=0$. 

Sun et al. \cite{sun_imaging_2008} found $\MTRasymss=PTR\cdot\alpha\cdot(1-\sigma)$ which combines the labeling efficiency $\alpha$ found by the weak-saturation-pulse approximation and  a spillover coefficient $\sigma$ from the strong-saturation-pulse approximation. This formula is only valid on resonance of pool \pb in contrast to Eq. \eqref{eqn:asymSS}.

Another approach, applicable for small \bi, eliminates the spillover effect by a probabilistic approach \cite{zaiss_quantitative_2011}.  
This Z-spectrum model taken from \cite{zaiss_quantitative_2011} yields
\begin{align}
\begin{split}
\text{PTR}(\dw)&\approx \frac{\Zssref -\Zsslab}{\Zssref -\Zsslab+\Zssref\cdot\Zsslab}\\
&=\frac{\Rexlab}{\cos^2\theta\cdot\R{1a}+\Rexlab} 
\end{split}
\end{align}
which turns out, after substitution of \Zss by Eq.\eqref{eqn:Zss}, to be an approximation of PTR if $\theta$ is small.
The asymmetry normalized by the reference scan was proposed for spillover correction \cite{liu_high-throughput_2010}.
Applying eq. \eqref{eqn:Zss} yields
\begin{align}
\label{eqn:reddyrefnorm}
\frac{\Zss(-\dw)-\Zsslab}{\Zssref} = \frac{\Rexlab}{\Reff(+\dw)+\Rexlab}
\end{align}
which again approximates PTR if $\theta$ is small.

By use of  Eqs. \eqref{eqn:Zss} and \eqref{eqn:Z_full_solution_apex} we obtain for the asymmetry in transient-state a bi-exponential function
\begin{equation}
\label{eqn:SOL_MTRasym}
\begin{split}
\MTRasym(\dw,t)=& \MTRasymss(\dw)\\
+e^{-\Rho(-\dw) t}\cdot & (\Pz\Pzeff-Z^{ss}(-\dw)) \\
-e^{-\Rho(+\dw) t}\cdot & (\Pz\Pzeff-Z^{ss}(+\dw)) .
\end{split}
\end{equation}
Neglecting direct saturation of pool \pa and assuming \Pz~=~\Pzeff~=~1 yields the mono-exponential approximation at the CEST resonance \cite{zhou_chemical_2006,mcmahon_quantifying_2006} 
 \begin{align}
\MTRasym(\dwb=0,t)=\MTRasymss(\dwb=0)\cdot(1-e^{-(\R{1a}+\ka)t}) ,
\end{align}
with the rate constant $\Rho = \R{1a} + \ka$.
This is valid if $\theta$ is small, leading to $\Reff\approx\R{1a}$ and, with the limit of Eq.~\eqref{eqn:SOL_Rexmax_ka}, $\Rex \approx \ka$ (solid red, Fig.~\ref{fig:Rex}b).

The ratiometric analysis approach QUESTRA  \cite{sun_simplified_2011} includes direct saturation and is independent of steady-state. It can be expressed by means of Eq.~\eqref{eqn:Zs} under the same assumptions  \Pz=\Pzeff=1 and $\Reff\approx\R{1a}$ and $\Rex\approx\ka$ 
\begin{align}
\text{QUESTRA}(t)=\frac{\Zs(+\dw,t)}{\Zs(-\dw,t)}\approx e^{-\ka t}.
\end{align}
\add{Another method, pCEST  \cite{vinogradov_pcest:_2012}, employs \Rho in an inversion recovery experiment.
The pCEST signal obeys the negative of Eq.~\eqref{eqn:SOL_MTRasym} if the initial inversion is introduced by 
$\Pzeff = -\cos\theta$. Hence, the full dynamics of the \Rho inversion recovery signal is }
\begin{align}
\add{\text{pCEST}(\dw,t)=-\MTRasym(\dw,t,\Pzeff = -\cos\theta)}
\end{align}
\add{The pCEST signal can be positive in transient-state, but is negative in steady-state.}
\add{ This inversion recovery approach was suggested first to increase SNR for MT effect by Mangia et al. \cite{mangia_magnetization_2011} and for SL already by Santyr et al. \cite{santyr_off-resonance_1994} and again by Jin and Kim  \cite{jin_quantitative_2012}. Their \textit{iSL} signal is in our notation equal to \Z(\dw,\wi,t) (Eq.\eqref{eqn:Z_full_solution_apex}) with \Pzeff=-1 and their projection factors for CEST and SL are identical with \Pz and \Pzeff. For \Rex the approximation of Ref. \cite{trott_r1rho_2002} is used, assuming $\R{2b}=\R{2a}$. Especially for the quantification employing different \bi their approach will benefit from our approximation of \Rex. By irradiation with Toggling Inversion Preparation (iTIP) Jin and Kim were able to remove \Zss which allows for direct exponential fit of the difference signal of SL and iSL and thus promises reduced scanning time \cite{jin_quantitative_2012}.}
 
\subsection{Separation for \Rex}
The dependence of CEST and SL on exchange is mediated by \Rex, the exchange-dependent relaxation rate in the rotating frame. Since the discussed evaluation algorithms for PTR depend on direct water saturation, we propose methods which use the underlying structure of the Z-spectrum and solve the solutions for \Rex.
For the transient state QUESTRA can be extended by inclusion of \Rex and the projection factors \add{(in \Zs, Eq. \eqref{eqn:Zs})} :
\begin{multline}
\label{eqn:ZZ}
\text{QUESTRA}_{R_{ex}}(t)=\frac{\Zs(+\dw,t)}{\Zs(-\dw,t)}= e^{-\Rex(\dw) t}.
\end{multline}
which provides direct access to \Rex. Even without creating \Zs one can measure the experimental \Rho(\dw) decay rate and obtains $\Rex(\dw) \approx \Rho(+\dw)-\Rho(-\dw)$ by asymmetry analysis of the rate \Rho(\dw).
For the evaluation of steady-state measurements we suggest an extension of Eq. \eqref{eqn:reddyrefnorm}
\begin{multline}
\label{eqn:MTnorm}
\text{MTR}_{R_{ex}}(+\dw)=\frac{\Zssref-\Zsslab}{\Zssref\cdot\Zsslab}  = \\
= \frac{1}{\Zsslab} - \frac{1}{\Zssref} = \frac{\Rexlab}{\cos\theta\cdot \Pz \cdot \R{1a}}
\end{multline}
which yields \Rex in units of \R{1a} and is independent of spillover. \Rex can be calculated by determination of \R{1a} and the projection factors . $\theta$ can be determined by  \bi mapping and \R{1a} can be measured, however \R{1a} is not the same as the observed relaxation rate \Robs in a inversion or saturation recovery experiment, especially if a macromolecular pool is present \cite{desmond_understanding_2012}.
Since MTR$_{R_{ex}}$ and QUESTRA$_{R_{ex}}$ evaluations employ directly Z-spectra data, they are useful saturation transfer evaluation methods for determination of \Rex with correction of direct saturation.
However, they are still asymmetry-based and are not applicable to systems with pools with opposed resonance frequencies. In this case, the most reliable evaluation is fitting whole Z-spectra  by  using Eq.\eqref{eqn:Z_full_solution_apex} including a superimposed \Rex of the contributing pools.

\subsection{Determination of \R{2b}, \kb and \fb}
As proposed by Jin et al. \cite{jin_spin-locking_2011} the width $\Gamma$ (Eq. \eqref{eqn:SOL_RexGamma}) of \Rex(\dwb) can be used to obtain \kb directly. But especially for small \kb the extension by \R{2b} is necessary. Fitting \Rexmax for different \bi yields \fb and \kb separately  similar to the QUESP method \cite{mcmahon_quantifying_2006} and Dixons Omega Plots \cite{dixon_concentration-independent_2010} , but again the neglect of \R{2b} in Eq. \eqref{eqn:SOL_RexmaxLS} will distort the values for \kb and \fb.
The width of \Rex is a linear function of $\wi^2$:
\begin{align}
\frac{\Gamma^2}{4}(\wi^2)= \frac{\kb+\R{2b}}{\kb}\cdot (\wi^2) + (\kb+\R{2b})^2
\end{align}
and 1/\Rexmax is a linear function of $\wi^{-2}$
\begin{align}
\frac{1}{\Rexmax (\wi^{-2})} = \frac{\kb+\R{2b}}{\fb}\cdot (\wi^{-2}) +  \frac{1}{\fb\kb}.
\end{align}
Hence, also the fit of Z-spectra for different \bi  yields \fb, \kb and \R{2b}, separately.

\section{Conclusion}
\label{seq:conc}
We extended the analytical solution of the \BM equations for SL by the relaxation rate \R{2b} and identified the projection factors necessary for application of the theory to CEST experiments. Temporal evolution as well as steady-state magnetization of \CEST and \SL experiments can be described by one single model governed by the smallest eigenvalue in modulus of the \BM equation system which is $-\Rho$. \Rho contains the exchange-dependent relaxation rate \Rex. We extended \Rex by the transversal relaxation \R{2b} which allows application of the theory to slow exchange, where \R{2b} is in the order of \kb and not negligible. \Rex of different pools can be superimposed to a multi-pool model even for a macromolecular MT pool. Compared to methods designed to estimate PTR, estimators of \Rex are less dependent on water proton relaxation. Finally, we showed that determination of \Rex as a function of \wi and \dw allows to determine concentration, exchange rate, and transverse relaxation of the exchanging pool.

\appendix
\section{\add{Eigenvector approximation}}
\label{seq:App_EV}
We consider the Taylor expansion in \fb of the eigenvector $\lambda_1$ of the smallest eigenvalue in modulus \li. The constant term of this expansion evaluated  on the resonance of pool \pb yields for the components of this eigenvector in pool \pa
\begin{align}
\label{eqn:APP_EVfull}
\begin{pmatrix}
\wi \\
0  \\
\dw
\end{pmatrix}
+
\underbrace{
\begin{pmatrix}
\frac{ (\R{1a}+\li)(\R{2a}+\li)}{\wi}  \\
\frac{\dw}{\wi} (\R{1a}+\li)  \\
0
\end{pmatrix} }_*
\end{align}
With the approximation of Eq. \eqref{eqn:reff} $\li=-\Reff=\R{1a}+(\R{2a}-\R{1a})\sin^2\theta= \R{2a}-(\R{2a}-\R{1a})\cos^2\theta$.
The first component of (*) can be neglected if
\begin{align}
|\frac{(\R{1a}-\Reff)(\R{2a}-\Reff)}{\wi}| \ll  \wi.
\end{align}
This yields
\begin{align}
(\R{2a}-\R{1a})^2 \ll \frac{\wi^2}{\sin^2\theta\cos^2\theta} 
						=  \frac{\wi^2}{\frac{\wi^2\dw^2}{\weff^4}} 
						=  \frac{\weff^4}{\dw^2} .
\end{align}
Since  $\frac{\weff^4}{\dw^2}>\weff^2$ this can be reduced to the condition
\begin{align}
\label{eqn:APP_EVcondition}
|\R{2a}-\R{1a}| \ll \weff.
\end{align}
The second component of (*) vanishes under the same condition \eqref{eqn:APP_EVcondition}.
After neglect of (*) and normalization the eigenvector of the smallest eigenvalue (Eq. \eqref{eqn:APP_EVfull}) simplifies to (Fig. \ref{fig:EVweff}a)
\begin{align}
\label{eqn:APP_EVapproximated}
\vec{v}_1=
\begin{pmatrix}
\sin\theta \\
0  \\
\cos\theta
\end{pmatrix}.
\end{align}
Along this eigenvector the Bloch-McConnell equations are one-dimensional
\begin{align}
\label{eqn:APP_stst_derivation}
\dot{M}_{z_{eff}}=\li \cdot M_{z_{eff}} + C_{z_{eff}}
\end{align}
where the constant part $C_{z_{eff}}$ is the projection of $\vec{C}$ (Eq. \eqref{eqn:BM_C}) on the eigenvector $\vec{v}_1$ \eqref{eqn:APP_EVapproximated} giving
\begin{align}
C_{z_{eff}}=\cos\theta\cdot\R{1a}.
\end{align}
The solution of Eq. \eqref{eqn:APP_stst_derivation} is the combination of the general solution of the homogeneous equation (which is an exponential function with rate \li) superimposed with a special solution of the inhomogeneous equation.
The steady-state is a special solution and is obtained by setting $\dot{M}_{z_{eff}}=0$ which gives
\begin{align}
-\li \cdot M_{z_{eff}} = \cos\theta\cdot\R{1a}.
\end{align}
By backprojection on the z-axis and normalization by \mo one obtains the steady-state solution Eq. \eqref{eqn:Zss}:
\begin{align}
\Zss = \frac{\Pz\R{1a}\cos\theta}{-\li}
\end{align}

\section{Eigenvalue approximation}
\label{seq:App}
The eigenvalues $\lambda_i $ of a $n\times n$-matrix are the roots of the normalized characteristic polynomial and are defined by
\begin{align}
\det(\A-\lambda\cdot\mathbf{I})=0\Leftrightarrow   \lambda^n + c_{n-1}\lambda^{n-1}+...+c_1 \lambda + c_0 =0
\end{align}
where
\begin{align}
c_0=(-1)^{n} \det(\A)=(-1)^{n} \lambda_1\cdot...\cdot\lambda_n
\end{align}
and
\begin{align}
c_1=(-1)^{n-1} \sum_{i=1}^{n}\frac{\lambda_1\cdot...\cdot\lambda_n}{\lambda_i}=-c_0\cdot\sum_{i=1}^{n}\frac{1}{\lambda_i}
\end{align}
The assumption that all eigenvalues are much larger than  
$|\lambda_1|\ll|\lambda_2|\leq...\leq|\lambda_n|$
leads to 
\begin{align}
c_1=-c_0\cdot(\frac{1}{\lambda_1}+\frac{1}{\lambda_2}+...+\frac{1}{\lambda_n})\approx -\frac{c_0}{\lambda_1}
\end{align}
This approximation is also valid for complex eigenvalues, because the conjugate complex is also an eigenvalue and therefore
$\frac{1}{\lambda_j}+\frac{1}{\lambda_j^*}=\frac{2Re(\lambda_j)}{|\lambda_j|^2}<\frac{2}{|\lambda_j|}$. Equations (A.2) and (A.4) allow general approximation of the smallest eigenvalue in modulus by
\begin{align}
\label{eqn:APP_c0c1}
\lambda_1=-\frac{c_0}{c_1}
\end{align}
The error is smaller than $|\frac{\lambda_1''}{\lambda_2''\cdot(n-1)}|$. Justified by linearization of the characteristic polynomial, expression \eqref{eqn:APP_c0c1} was also suggested in Ref. \cite{trott_r1rho_2002}.


\bibliographystyle{model1-num-names}
\bibliography{zotero} 


\end{document}